\documentclass[%
 aip,
cp,  
 amsmath,amssymb,
 reprint,%
]{revtex4-2}

\usepackage{graphicx}
\usepackage{dcolumn}
\usepackage{bm}

\usepackage[utf8]{inputenc}
\usepackage[T1]{fontenc}
\usepackage{mathptmx} 
\usepackage{float}
\usepackage{hyperref}

\begin{document}

\title{Social Reader Perusall -- a Highly Effective Tool and Source\\
of Formative Assessment Data
}

\author{Jozef Han\v{c}} 
 \email[\textit{corresponding author:} ]{jozef.hanc@upjs.sk}
\affiliation{
  Institute of Physics, Faculty of Science, P. J. Šafárik University in Košice, Slovakia 
}
\author{Martina Han\v{c}ov\'a}
 \email{martina.hancova@upjs.sk}
\affiliation{%
Institute of Mathematics, Faculty of Science, P. J. Šafárik University in Košice, Slovakia 
}%

\author{Dominik Borovsk\'y$^{1,}$}
 \email{dominik.borovsky@student.upjs.sk}

\date{\today} 

\begin{abstract}
The contribution provides a detailed exploration of the online platform Perusall as an advanced social annotation technology in teaching and learning STEM disciplines. This exploration is based on the authors' insights and experiences from three years of implementing Perusall at P.J. Šafárik University in Košice, Slovakia. While the concept of social annotation technology and its educational applications are not novel, Perusall's advanced features, including AI and data science reports, enable its use in both synchronous and asynchronous blended and flipped learning environments. In this context, Perusall serves as a digital tool for collecting formative data, monitoring student progress, and identifying areas of difficulty. This assessment data can be effectively utilized in preparing and personalizing subsequent face-to-face group interactions, thereby enhancing and improving the learning experience. From a pedagogical viewpoint, Perusall's role was particularly significant during the Covid-19 pandemic, enabling effective, continuous, and engaging learning amidst social distancing and physical restrictions. Today, Perusall has become a key tool in blended learning, facilitating higher-order cognitive processes during the educational process and, with its multifaceted applications, serves as a modern catalyst in redefining educational experiences and outcomes.
\end{abstract}

\maketitle

\section{\label{sec:level1}INTRODUCTION: PERUSALL EXCHANGE 2022}

In the second half of May 2022, researchers, educators, and teachers had the opportunity to participate in an unconventional online event – the asynchronous social conference \textit{Perusall Exchange 2022} \cite{mazur_exchange_2022,mazur_2022_2022}. This annual conference lasted 12 days (May 16th – May 27th), and the organizers of the conference were none other than the creators of Perusall from Harvard University, led by the renowned physicist and physics educator Professor Eric Mazur. The central theme of the second edition was \textit{Social Learning}, and one of its tracks was \textit{Peer Instruction 30}, in honor of the 30th anniversary of the development of peer instruction by Eric Mazur and his collaborators - one of the most successful interactive teaching methods applied not only in STEM subjects but also in humanities-oriented subjects \cite{crouch_peer_2001,simkins_just_time_2009}. The conference became unique in its format as it was entirely asynchronous, taking place in a virtual digital space, particularly in the online platform known as the social e-reader  Perusall (\url{https://www.perusall.com/}). 

The creators of this freely available and cost-free digital platform, established in 2016, were the very conference organizers from Harvard – G. King, B. Lukoff, E. Mazur, and K. Miller \cite{miller_use_2018}. Perusall, as digital technology, is classified as a social e-reader and allows synchronous but mainly asynchronous and jointly active viewing, studying, and learning through individual annotations of any digital content or digital medium. Throughout the duration of the aforementioned conference, participants (and also the authors of this article), who registered in the Perusall conference, could engage asynchronously in group discussions on the posted contributions at any time. 

More than 50 contributions from keynote speakers, panelists, and presenters took various digital forms (Fig. 1, top) – podcast format (5 contributions), video presentations (32 contributions), short articles (12 contributions), posters (8 contributions), and live event recordings (8 contributions). Overall, more than 2000 participants engaged in discussions on individual contributions.

At the bottom of Fig. 1, we see a specific poster from this conference \cite{sun_collaborative_2022}, where a participant can annotate the given electronic resource, i.e., highlight any part of the text and write a comment or question in relation to it. This action creates a separate conversation thread for each annotation, where a discussion can take place among the participants. As an interesting fact, we can mention that among the five most annotated or active presentations \cite{mazur_2022_2022} was a video contribution by Eric Mazur himself, presenting and discussing the current state and innovations in the Peer Instruction method after 30 years of its existence \cite{mazur_new_2022}.

\begin{figure}
    \centering
    \includegraphics[width=1\linewidth]{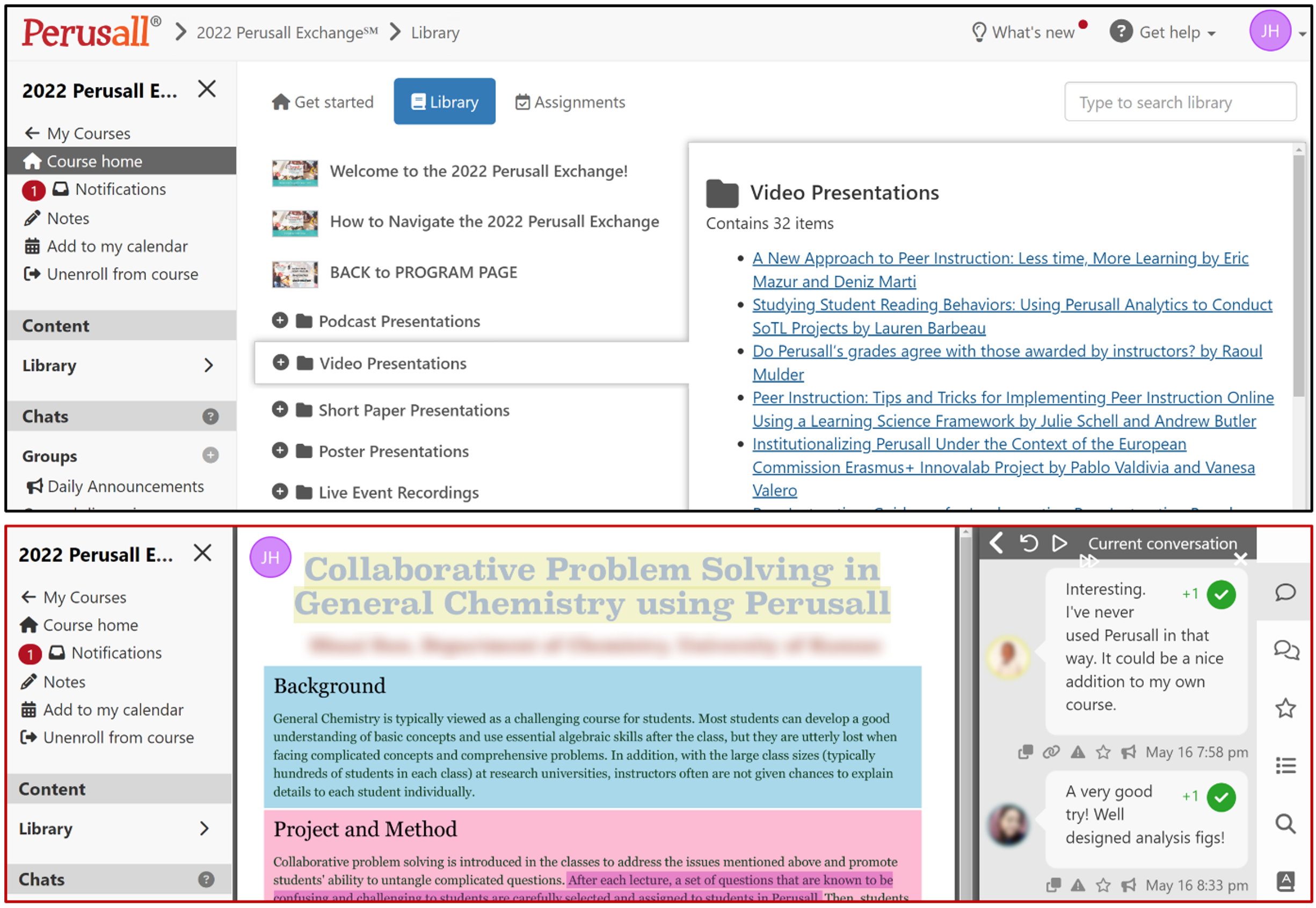}
    \caption{A view of the online platform Perusall during the Perusall Exchange 2022 conference. At the top, we see the types of contributions \cite{mazur_exchange_2022}, and at the bottom, we see a conference contribution in the form of a poster \cite{sun_collaborative_2022}.}
    \label{fig:1}
\end{figure}

In 2023, this exceptionally successful conference dedicated to social learning will again be organized in June, with key topics being Artificial intelligence (AI) and Deep Learning Technologies in Education, Learning Communities, Alternative Grading, and Assessment Approaches. The main objective of this article is to provide a detailed introduction and explanation of what social annotation, asynchronous learning mode, and social e-reader mean in the context of social learning, how Perusall operates, and its potential educational applications. Finally, we would like to share our over three-year experiences with Perusall, which we gained during the years 2020 – 2022 at the P.J. Šafárik University in Košice, Slovakia.

\section{OUR DIRECT EXPERIENCE WITH PERUSALL}
The concept of social annotation technology (or social e-readers) as a learning or educational promotion tool is not novel. According to \cite{novak_educational_2012}, in 2012 this technology has existed for at least the past 10 years. Its essential feature, as online social bookmarking, allows users to annotate, i.e., to highlight specific text or any part of an electronic source (podcast, video, webpage, pdf, word) adding their own comments and questions.  At the same time, it enables the easy sharing of these annotations, promoting chats, discussions and interactions among the participants. 

From the technological viewpoint, today we know several social annotation tools, such as Diigo (\url{https://www.diigo.com/}), Google Docs (\url{https://docs.google.com/}), or Kami (\url{https://www.kamiapp.com/}). An example of a more advanced and sophisticated platforms is just Perusall, which we apply in education and aim to describe in this section in more detail. However, there are more or less equivalent alternatives to Perusall, such as open-source Hypothesis (\url{https://web.hypothes.is/}, \cite{kalir_value_2022} or NotaBene, shortly NB (\url{https://nb.mit.edu/}, \cite{zyto_successful_2012}) which was a probably inspiration for Perusall \cite{miller_analysis_2016}.

\subsection{Perusall as a modern digital educational technology}

Perusall, as one of the most recent and advanced social annotation tools, is developed and based on extensive educational data analytics and insights from behavioral and educational research. It was primarily created to motivate and engage students, with the intention that these students look forward to working and studying within it. To summarize and mainly to understand the didactic significance of Perusall, it is important to note that Perusall, as a social e-reader, is a combination of three digital technologies (see Fig. 2). Perusall simultaneously represents:

\begin{itemize}
    \item \textbf{Social Network:} a given group of students or a community of participants is interconnected in a private digital network, with each annotation opening a new conversational thread allowing interaction much like chatting with friends on social networks (Facebook, Twitter, Instagram).
    \item \textbf{Annotation Tool:} an instructor or student can easily highlight part of a text, image, or specific moment in a video or podcast, to which they can write their comments or questions. Perusall allows writing and formatting text like a text editor (e.g., Word), and comments can include multimedia (e.g., images and videos), emoticons, equations in LaTeX, codes of various programming languages, or even e-voting questions as in response systems. Important points can be highlighted with hashtags; to be not distracted, Perusall also offers the option to turn off all annotations or to display only a certain type of them (instructor's, mine, selected group's, or individual's).
    \item \textbf{Data Science Report Tool:} the instructor receives automated, detailed, comprehensive, ongoing, and overall reports of all data from students’ learning; it includes basic descriptive statistics (e.g., number of posts, their form, student activity over time, on individual pages of a document, distribution of grades), but also elements of AI and data science, where it automatically grades or via machine learning automatically assesses the quality of participants' contributions (three basic levels of quality 0, 1, 2).
\end{itemize}

\begin{figure}
    \centering
    \includegraphics[width=1\linewidth]{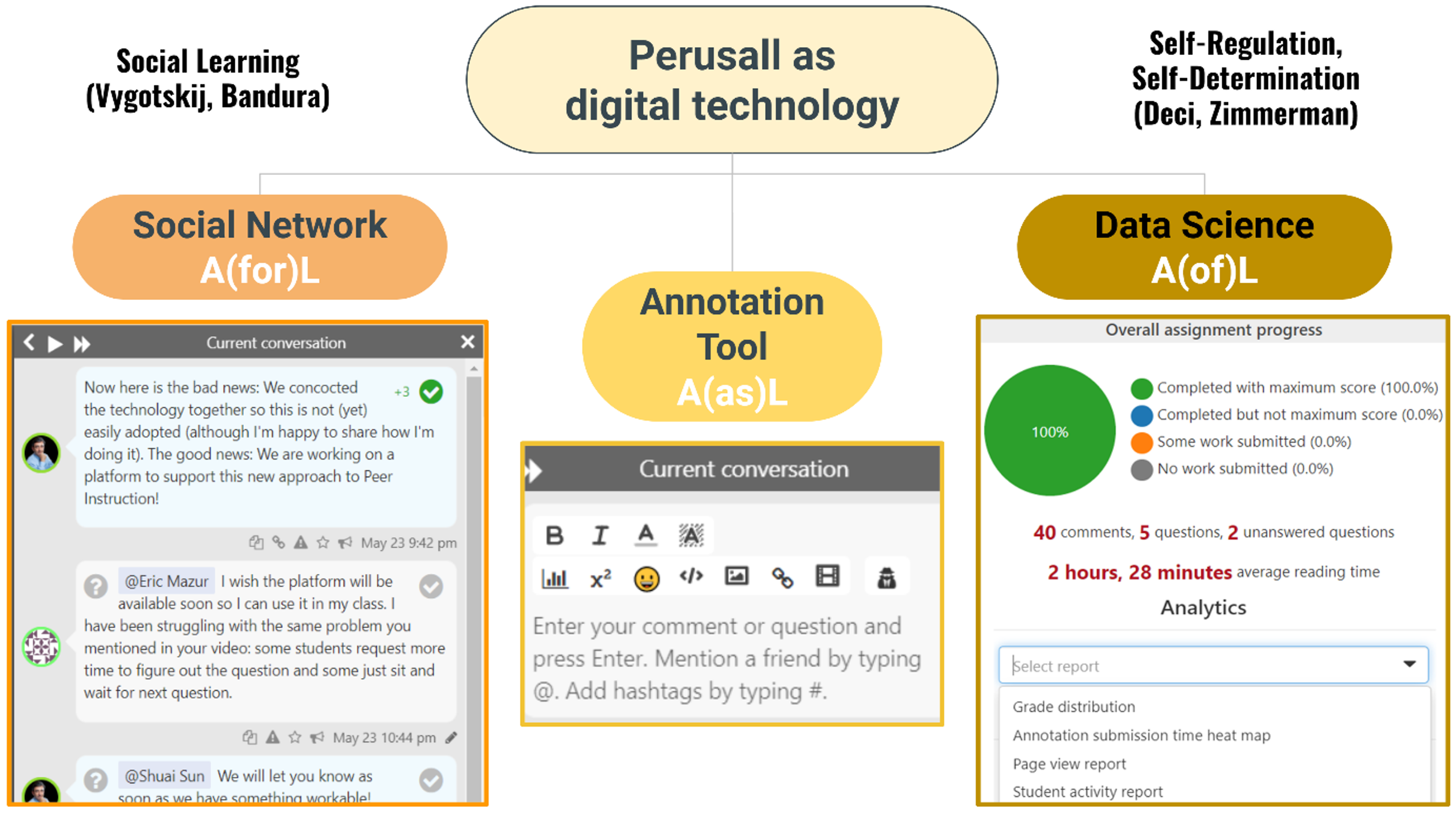}
    \caption{Perusall as a modern digital educational technology combining three digital technologies.}
    \label{fig:2}
\end{figure}

Perusall can be integrated into well-known LMS like Canvas, BlackBoard, or Moodle, making it easy for instructors to incorporate into their courses and for students to access the platform.

Regarding a psychological and social viewpoint, students do not study in a virtual space alone but collaboratively and interactively, i.e., Perusall naturally incorporates elements of social learning (Vygotsky, Bandura). Social learning was very significant from our perspective during the COVID pandemic and social distancing when students felt loneliness and isolation. Furthermore, the asynchronous mode of study and interactions allows self-regulation and self-determination (Deci, Zimmerman) in students’ learning behavior and also suppresses the undesirable intellectual dependency of students on instructors.

From a didactic viewpoint, Perusall also represents the application of one of the newest views on assessing the educational process as a proven important factor of effective education. According to Earl \cite{earl_assessment_2013}, the assessment can be seen as an \textit{assessment OF/FOR/AS learning}. From this perspective:

\begin{itemize}
\item \textbf{AI and Data Science analytics} (via data reports) represent the first component of assessment -- \textit{assessment OF learning}; it is nothing else as a summative assessment, which is, however comprehensive and automated, providing a detailed picture of the given data showing progress, but also problems, uncertainties, misconceptions of students. Since data analytics is immediately accessible at any phase of students learning, this allows intervention in a formative sense. Time management problems, when or at the last minute work, can also be addressed.
\item \textbf{Social learning and interaction} (via social network) provide opportunities for assessment, which we refer to as \textit{assessment FOR learning}; it is a lower level of formative assessment, where feedback and control occur between the teacher and students or students themselves.
\item \textbf{Self-regulated and self-determined learning behavior} (via own annotations using annotation tool) develop the assessment referred to as \textit{assessment AS learning}, a higher form of formative assessment, where the student himself takes responsibility for his own learning, allowing to study at own pace, fuller attention, concentration, honesty, and self-discipline. 
\end{itemize}

\subsection{Perusall as a blended learning tool}

As for direct application in education, Perusall now belongs to the key educational technologies used in blended learning and its version known as flipped learning \cite{miller_use_2018,talbert_flipped_2017,tucker_blended_2017}. In our teaching practice, we utilize an interdisciplinary, modern active learning STEM approach within the blended learning framework in over 10 courses dealing with STEM subjects at our university \cite{gajdos_interactive_2022,hanc_open_2022,pankova_practical_2017}. Since the pedagogical details of our approach can be found in the mentioned references, we provide two illustrative examples from our recent courses – Modern Didatical Technologies (MDT), in 2022 enrolled by a small group of students ($N = 4$), and Physics Practical I (PPI), in 2021 enrolled by 10 students.

The primary aim of the MDT is that a future pre-service teacher should acquire key digital skills and competencies to be more effective in school practice. In Fig. 3, we see a screenshot of the Perusall platform for the course with assignments, e.g., the first assignment deals with the concept of the modern hybrid classroom of the 21st century. In accordance with the flipped learning approach, prior to a face-to-face group session, students asynchronously but collaboratively learn about the topic from a digital study source via annotation activities.

During this first flipped learning phase, Perusall continuously collects formative data monitoring students' progress as well as their difficulties. The Perusall data report captured from this assignment, seen in Fig. 2 on the right, shows that 4 students made 40 comments, posed 5 questions, and completed the Perusall assignment with a maximum score during an average reading time of 2 and a half hours. After the phase, these Perusall data are used to prepare, adjust, and personalize the subsequent face-to-face group interaction, taking into account the student’s specific needs and levels. Perusall allows us to teach in a way that focuses rather on the most important than the non-problematic areas.

\begin{figure}
    \centering
    \includegraphics[width=1\linewidth]{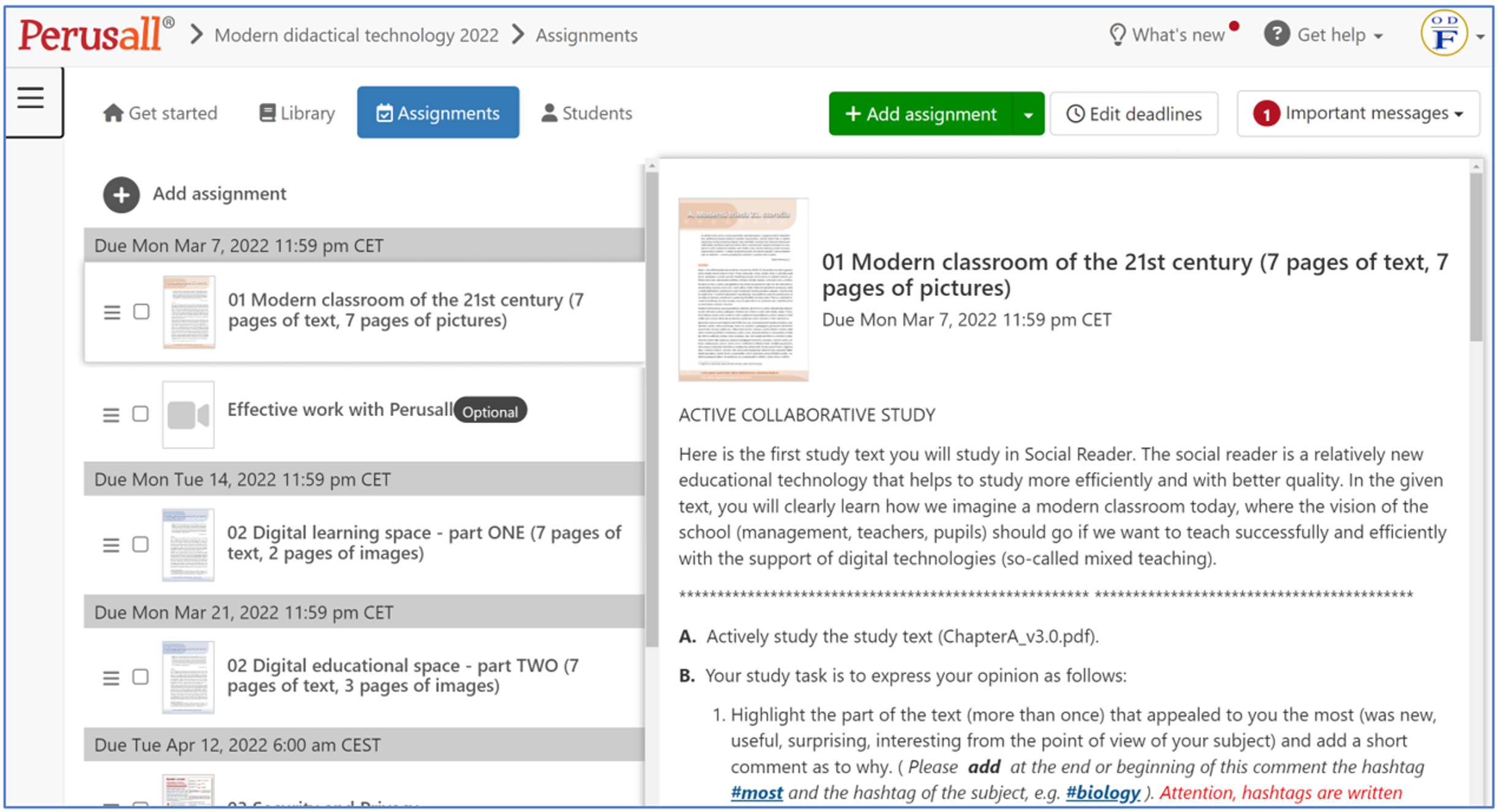}
    \caption{A screenshot from our Perusall course, Modern Didactical Technologies 2022,intended for STEM and humanities students at P.J. Šafárik University}
    \label{fig:3}
\end{figure}

Regarding the course PPI, we also use Perusall in a synchronous mode, during which, in the face-to-face session of blended learning, students are divided into groups (2-3 students per group), with each group working on one of the stations according to the station rotation model \cite{tucker_blended_2017}. Students rotate through learning stations in the classroom, and one of the stations involves a Perusall group study of a video tutorial focusing on an important step of the measurement process. 

Our three-year pedagogical experience with Perusall, during which we implemented a total of 24 courses in STEM disciplines with more than 500 students ($N=502$), confirmed the general findings of previous educational research \cite{miller_use_2018, porter_collaborative_2022}. Indeed, typically over 90\% of students engaged prior to face-to-face flipped learning sessions with Perusall (only in three courses we had an engagement of less than 90\%, between 80-90\%). In small groups of up to 10 students, this was always nearly 100\%. Based on observations and results of midterm and final exams, using Perusall led to better, more concentrated students’ work during face-to-face sessions, allowing more time for tasks involving higher cognitive processes and real-world context. It also resulted in improved learning engagement, focused discussion, increased learning achievement, and better reading comprehension ability.

\subsection{Data science automatization in processing Perusall assessment data}
In blended learning with Perusall, the educational researcher or teacher (as well as the student) gains much more educational data than traditional teaching and learning. Perusall standardly provides the teacher with a set of data reports, which can be used mainly for formative assessment. However, manipulating these reports often requires many clicks, and working in multiple courses at once can be irritating and very time-consuming. If we want specific data and its processing, manually downloading this data from Perusall and subsequently processing it, e.g., in Excel, is not easily reproducible, is still very tedious and painful, and there is a significant risk of errors.

Therefore, we decided to apply Open Data Science Tools to work with data from Perusall for educational purposes, where the leading technology is Jupyter. A concrete example of such data science output using modern data visualization and summarization tools in the form of a Jupyter notebook is shown in Fig. 4, where we see part of the summary of students' work ($N=4$) in the course MDT 2022 concerning the number, length, and quality of comments and questions within individual assignments as well as overall. Here, it is sufficient to know that open data science tools allow for the automation of any manipulation of data, or their processing and analysis. More detailed information on what these open tools are and the Jupyter technology can be found in our articles \cite{gajdos_interactive_2022, hanc_open_2022}.

\begin{figure}
    \centering
    \includegraphics[width=0.8\linewidth]{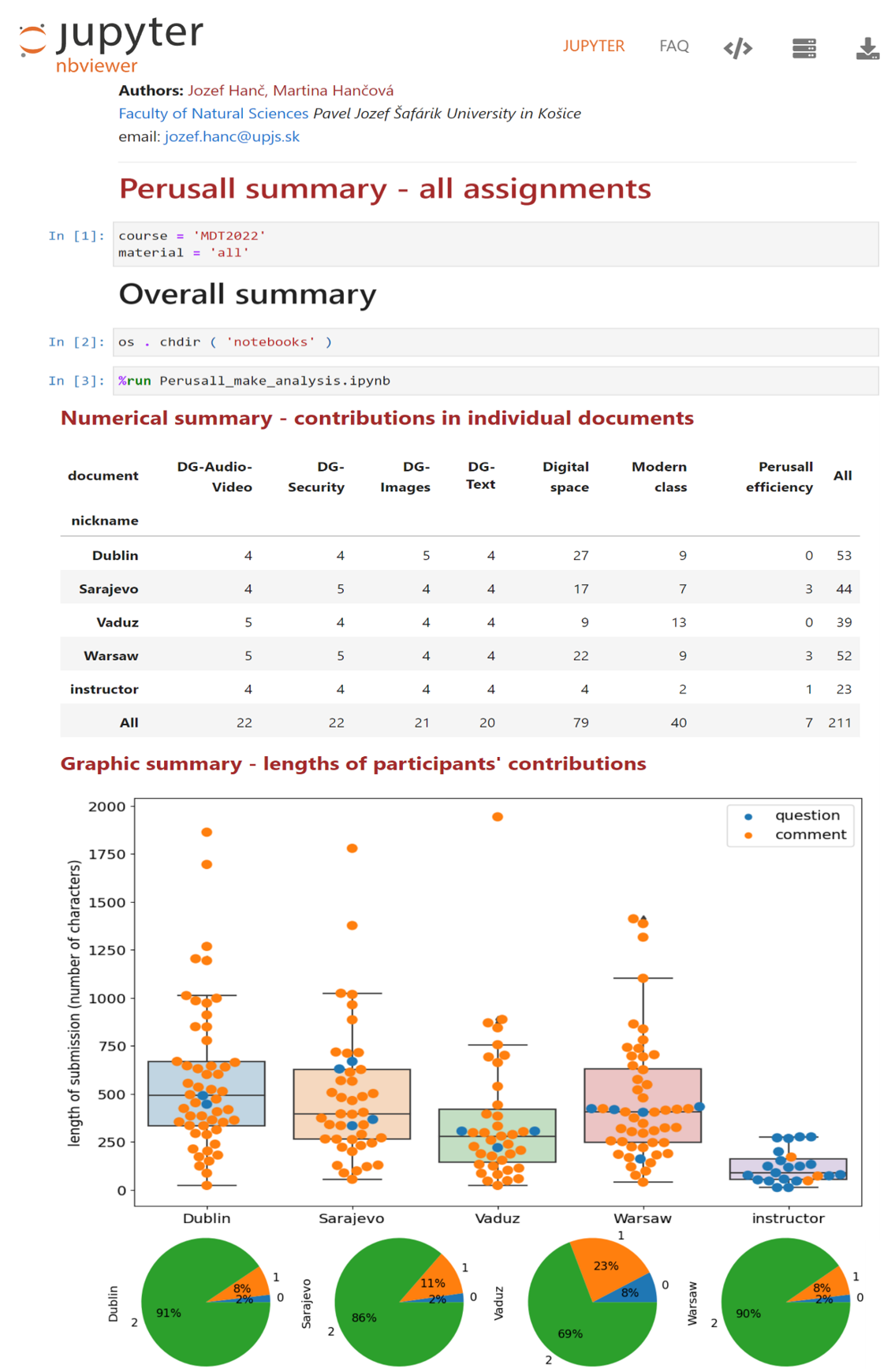}
    \caption{A screenshot from our Perusall course, Modern Didactical Technologies 2022,intended for STEM and humanities students at P.J. Šafárik University}
    \label{fig:4}
\end{figure}

Regarding Fig. 4, we highlight data visualization through the combined use of boxplots and swarmplots. Boxplots summarize key statistical measures, such as the median, quartiles, and potential outliers that lie beyond the whiskers (more than 1.5 times the interquartile range). On the other hand, swarmplots are categorical scatter plots that visualize the distribution of values where each data point is plotted as a dot, and the dots are spread out horizontally to avoid overlap. Two different colors (blue and orange) represent two categories of participants’ posts – comments and questions. We employ a similar Jupyter notebook for data analysis during any assignment, where in just a few milliseconds, the teacher can see various aspects such as students' activity, the most active discussions, problematic comments, or comments sorted by hashtags. Automating data manipulation significantly reduces time and effort, effectively aiding in preparing face-to-face sessions.

In creating these notebooks, we utilized two open data science tools – Python Selenium and the Pandas library. Python Selenium can be thought of as a virtual robotic arm that understands Python commands. It can autonomously open Perusall and download the data we need in the xlsx format. Subsequently, the Python Pandas library can be used for data processing. In Pandas, you can perform individual steps of processing akin to Excel but using commands, obtaining processed results within milliseconds. 

From a data science and research perspective, we have been scientifically addressing the issue of efficient processing, analysis, and modeling of data from measurements in various fields of human activity for several years. The measurement in didactics and the use of open data tools were first addressed in the doctoral thesis of one of our former students, now a data analyst \cite{strauch_measuring_2020}. This issue is among the applications of advanced mathematical and statistical methods using open data science tools, which are central to our current data science scientific projects: \textit{Advanced Mathematical and Statistical Methods for Measurement} (APVV-21-2016) and \textit{Optimal Decision-making and Control Methods in Complex Data }(APVV-21-0369).

Finally, it is worth mentioning that one of us (DB), as a student, became a scientific research assistant in the last two years of his master’s studies in support of these projects (Topic: Application of data science tools in efficient automation and reproducibility of scientific work in educational research), and from the new academic year, will be pursuing a Ph.D. in Physics Education Research. His scientific work will also be related to research in the application and testing of modern statistical methods and diagnostic tools on data from educational experiments.

\section{Conclusion}
Perusall has emerged as a key tool in the educational landscape, particularly in blended and flipped learning environments. By encouraging active engagement with learning materials, Perusall frees up more time for higher-order cognitive processes, as outlined in Bloom’s Taxonomy, during face-to-face sessions. Perusall can have a pivotal role in preparation for science labs across STEM disciplines like Physics, Chemistry, and Biology by enabling students to come prepared and with a deeper understanding of the subject matter. Concerning our teaching during the Covid-19 pandemic, Perusall played a crucial role as a lifesaver, enabling effective, continuous, engaging learning despite social distancing and physical restrictions.

As for other educational applications (as were introduced e.g., in Perusall Exchange 2022 conference), Perusall enables a novel pedagogical design in homework assignments, including those in traditional paper-and-pencil formats. Through features that support social, collaborative learning, Perusall enhances problem-solving exercises and fosters a range of learning forms and methods, such as Peer Instruction. It proves to be invaluable in workshops, seminars, projects, and team meetings by providing a platform for seamless collaboration and interaction. Moreover, Perusall can improve continuous teaching practices in preparing future STEM teachers. It can also be effectively used in preparing and training high school students for scientific competitions. Perusall, with its multifaceted applications, serves as a modern catalyst in redefining educational experiences and outcomes.

\begin{acknowledgments}
This work is supported by the Slovak Research and Development Agency under the Contract no. APVV-21-0216 and APVV-21-0369.
\end{acknowledgments}

\bibliography{References}

\end{document}